\documentclass{elsart}
\usepackage[dvips]{graphicx}
\journal{Physica A}
\begin{document}
\begin{frontmatter}
\title{Multiple Shocks in a Driven Diffusive System with Two Species of Particles} %
\author{Farhad H Jafarpour}
\address{Physics Department, Bu-Ali Sina University, Hamadan,Iran\\
Institute for Studies in Theoretical Physics and Mathematics
(IPM), P.O. Box 19395-5531, Tehran, Iran}
\ead{farhad@ipm.ir}
\begin{abstract}
A one-dimensional driven diffusive system with two types of
particles and nearest neighbors interactions has been considered
on a finite lattice with open boundaries. The particles can enter
and leave the system from both ends of the lattice and there is
also a probability for converting the particle type at the
boundaries. We will show that on a special manifold in the
parameters space multiple shocks evolve in the system for both
species of particles which perform continuous time random walks on
the lattice.
\end{abstract}
\end{frontmatter}

\section{Introduction}
\label{Sec1}
Recently much attention has been paid to the study of
one-dimensional driven diffusive systems because these relatively
simple systems show a variety of critical phenomena such as
out-of-equilibrium phase transitions and spontaneous symmetry
breaking \cite{sch,sz}. Another remarkable feature of these
systems is that under special circumstances, for example by
imposing restrictions on microscopic reaction rates, shock waves
evolve in the system in the steady state \cite{fer}. A simple
model of this kind is the Totally Asymmetric Simple Exclusion
Process (TASEP). The TASEP is a system of particles with hardcore
exclusion interactions which hop only in one direction
on a one-dimensional lattice with open boundaries. This diffusion
model can describe hopping conductivity in ionic conductors,
traffic flow and interface growth \cite{sch,sz}. The TASEP shows
rich non-equilibrium behaviors such as shock wave \cite{djl,dls,dgls}, 
boundary induced phase transition \cite{kr}, and dynamical scaling 
in the universality class of the Kardar-Parisi-Zhang equation \cite{gs,ki}. 
This model has been solved exactly using a so-called Matrix Product Formalism (MPF)
\cite{dehp}. According to this formalism the stationary state
probability distribution function of the system can be written in
terms of expectation values of products of two non-commuting
matrices which are associated with the existence of
particles and holes at each site of the lattice and satisfy a 
quadratic algebra.\\
Another model of this kind is Partially Asymmetric Simple
Exclusion Process (PASEP) with open boundaries. In this model,
which is defined on an integer lattice $Z$ of length $L$,
each site of the lattice $i$ ($1 \leq i \leq L$) is either empty
$\tau_i=0$ or occupied by at most one particle $\tau_i=1$. The
system evolves according to a stochastic dynamical rule: during
each infinitesimal time step $dt$ the transitions allowed for the
bond ($i,i+1$) are $(1 0 \rightarrow 0 1)$ with rate $1$ and $(0 1
\rightarrow 1 0)$ with rate $x$. The parameter $x$ is positive and
measures the strength of the driving field. Without losing
generality one can assume that $x < 1$. Particles are injected
from the first and the last sites of the lattice with rates
$(1-x)\alpha$ and $(1-x)\delta$. They also leave the lattice from
both the first and last sites with rates $(1-x)\gamma$ and
$(1-x)\beta$ respectively. It is known that the phase diagram of
the PASEP has three different phases depending on $\alpha, \beta,
\gamma, \delta$ and $x$: in the steady state it has a high-density
phase, a low-density phase, and a maximal-current phase. Recent
investigations show that for the PASEP with open boundaries (and
even on an infinite lattice) a travelling shock with a step-like
density profile might evolve in the system provided that
microscopic reaction rates are tuned appropriately \cite{bs,kjs}.
The density of particles on the left and the right hand sides of
the shock position is a function of these rates. The shock
position then performs a random walk in the bulk of the system and
also reflects from the boundaries. There is also a possibility for
the existence of $n$ consecutive shocks in the system. In an
infinite system consecutive multiple shocks evolve according to
$n$-particle dynamics \cite{bs}. By slightly different definition
of the shock it has been shown that the same phenomenon takes
place for the PASEP on a finite lattice with open boundaries
\cite{kjs}. The PASEP has also been studied using the MPF
\cite{san,sasa}. It has been shown that its quadratic algebra has
an $n$-dimensional matrix representation \cite{er,ms} provided
that exactly the same constraints necessary for the existence of
$n$ consecutive shocks in the system hold \cite{kjs}. \\
Despite the remarkable work on the dynamics and structure of
shocks in the systems with one species of particles, not much is
known about the shocks in the systems with more than one species
of particles. In \cite{rs} the authors have introduced a
one-dimensional driven lattice gas with two types of particles and
nearest neighbor hopping. Part of their work is devoted to the
study of a single shock dynamics in the system. They have
investigated the time evolution equation of a product shock
measure and shown that the shock is stable and performs a random
walk provided that some constraints are satisfied. In this paper
we study the same model; however, we investigate the possibilities
for the existence of multiple shocks in the system. It turns out
that by imposing some conditions on the microscopic reaction
rates, stable multiple shocks evolve in the system. For the
special case where only a single shock exists, our results
converge to the ones obtained in \cite{rs}. We will also show that
under these constraints the quadratic algebra of this model can 
be mapped to the one which appears for the PASEP.\\
This paper is organized as follows. In section \ref{Sec2} we will
briefly review the PASEP from the MPF point of view.
In section \ref{Sec3} we will define the model and apply
the MPF to study its steady state properties. In section
\ref{Sec4} the conditions for the existence of multiple shocks in
our model will be studied. The concluding remarks will be given in
the last section.
\section{The PASEP}
\label{Sec2}
\begin{figure}
\centering
\includegraphics[height=3.5cm]{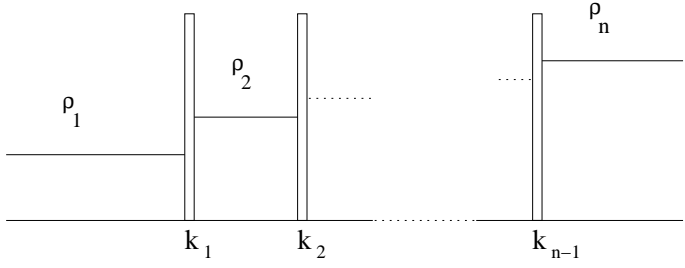}
\caption{\label{fig1}A typical Bernoulli shock measure for $n-1$
consecutive shocks for the PASEP. The density of particles at the
shock positions $k_i$ is 1.}
\end{figure}

In this section we will first review the steady state properties
of the PASEP with open boundaries defined in section \ref{Sec1},
because as we will see the steady state properties of our
two-species model are closely related to those of the PASEP. The
exact steady state properties of the PASEP for all values of the
boundary and bulk parameters might be calculated by using the MPF
\cite{sasa} which involves the representation theory of a
quadratic algebra equivalent to a $q$-deformed harmonic oscillator
algebra \cite{san}. According to the MPF the stationary
probability distribution function of the PASEP is given by
\begin{equation}
P( \{ \tau_1,\cdots,\tau_L \} ) \propto \langle W \vert
\prod_{i=1}^{L}(\tau_i D + (1-\tau_i) E)\vert V \rangle
\end{equation}
provided that the operators $D$ and $E$ besides the vectors
$\langle W \vert$ and $\vert V \rangle$ satisfy the quadratic
algebra \cite{ms}
\begin{eqnarray}
& & DE-xED=(1-x)(D+E) \label{A1a} \\
& & (\beta D - \delta E) \vert V \rangle = \vert V \rangle \label{A1b} \\
& & \langle W \vert (\alpha E - \gamma D) = \langle W \vert. \label{A1c}
\end{eqnarray}
The non-commuting operators $D$ and $E$ stand for the existence of
a particle and a hole at each site of the lattice. It has been
shown that the associated quadratic algebra (\ref{A1a}-\ref{A1c}) has exactly
one $n$-dimensional irreducible representation for any finite $n$
provided that the following constraint is satisfied by the bulk
and the boundary rates \cite{er,ms}
\begin{equation}
\label{C1}
x^{1-n}=\kappa_{+}(\alpha,\gamma)\kappa_{+}(\beta,\delta)
\end{equation}
in which we have defined
\begin{equation}
\kappa_{+}(u,v)=\frac{-u+v+1+\sqrt{(u-v-1)^2+4uv}}{2u}.
\end{equation}
Typical $n$-dimensional representations for (\ref{A1a}-\ref{A1c}) are given
in \cite{er} and \cite{ms}. The finite dimensional Fock representations allow 
us to drive exact results for the PASEP on some special curves of the phase 
diagram. Theses curves pass through the high-density and the low-density phases
but not the maximal-current phase. \\
The evolution of product shock measures in the PASEP has been
studied both for open boundary condition and on an infinite
lattice. In the following we study the dynamics of travelling
shocks in the PASEP with open boundaries briefly. We define a
Bernoulli measure for $n-1$ consecutive shocks as a product
measure with density $1$ at the shock positions $k_i$ and
intermediate densities $\rho_i$ between sites $k_{i-1}$ and $k_i$.
This product measure can be written as follows
\begin{equation}
\begin{array}{c}
\label{BSM} 
\vert k_1,k_2,\cdots,k_{n-1} \rangle =  \\   \vert \rho_1
\rangle^{\otimes (k_1-1)} \otimes \vert 1 \rangle \otimes \vert
\rho_2 \rangle^{\otimes (k_2-k_1-1)} \otimes  \vert 1
\rangle\otimes \cdots \otimes \vert 1 \rangle \otimes  \vert
\rho_n \rangle^{\otimes (L-k_{n-1})}.
\end{array}
\end{equation}
in which $0 < \rho_1 < \rho_2 < \cdots < \rho_{n} < 1 $. A typical
shock measure is plotted in Fig.~\ref{fig1}.
Now the question is that how (\ref{BSM}), as an initial
configuration, evolves in time. This is given by the master equation.
It is known \cite{bs,kjs} that for the PASEP with open boundaries
the time evolution equation of the Bernoulli shock measure (\ref{BSM})
has a closed form and is similar to the time evolution equation of
$n-1$ random walkers on a finite lattice provided that we have
\begin{eqnarray}
\rho_1=\frac{1}{1+\kappa_{+}(\alpha,\gamma)} \label{C2} \\
\rho_{n}=\frac{\kappa_{+}(\beta,\delta)}{1+\kappa_{+}(\beta,\delta)}
\label{C3}
\end{eqnarray}
and that the consecutive densities are related through
\begin{equation}
\label{C4}
\frac{\rho_{i+1}(1-\rho_i)}{\rho_i(1-\rho_{i+1})}=\frac{1}{x}
\end{equation}
in which $i=1,\cdots,n-1$. By iteration (\ref{C4}) leads to
\begin{equation}
\label{C5} \frac{\rho_n(1-\rho_1)}{\rho_1(1-\rho_n)}=x^{1-n}.
\end{equation}
The $i$th random walker hops to the left and the right with rates
$L_i=\frac{1-\rho_i}{1-\rho_{i+1}}x$ and
$R_i=\frac{1-\rho_{i+1}}{1-\rho_i}$ respectively.
Using (\ref{C2}) and (\ref{C3}) one can rewrite (\ref{C5}) in
terms of $\kappa_{+}(\alpha,\gamma)$ and
$\kappa_{+}(\beta,\delta)$ to see that it is simply the condition 
(\ref{C1}) i.e. the necessary condition for the existence of an 
$n$-dimensional representation for (\ref{A1a}-\ref{A1c}). One can now
construct the stationary state of the PASEP in terms of a linear
superposition of Bernoulli shock measures with $n-1$ shocks;
therefore, the $n$-dimensional representation of the quadratic
algebra describes the stationary linear combination of shock
measures with $n-1$ consecutive shocks.
\section{Two-species model with open boundaries}
\label{Sec3}
The model is defined on a finite lattice of length $L$ with two
species of particles. Each site of the lattice is either empty or
occupied by a particle of type $A$ or $B$. We assume that in
the bulk of the lattice the particles hop to the left and right
according to the following reaction rules
\begin{equation}
\label{A} \mbox{Bulk:}
\left\{%
\begin{array}{ll}
    A 0 \rightarrow 0 A & \hbox{\mbox{with rate 1}} \\
    0 A \rightarrow A 0 & \hbox{\mbox{with rate x}} \\
    B 0 \rightarrow 0 B & \hbox{\mbox{with rate 1}} \\
    0 B \rightarrow B 0 & \hbox{\mbox{with rate x}} \\
    A B \rightarrow B A & \hbox{\mbox{with rate y}} \\
    B A \rightarrow A B & \hbox{\mbox{with rate y}}. \\
\end{array}
\right.
\end{equation}
At the boundaries the particles are injected and extracted. The
particle type can also be converted there. The boundary processes
are defined as follows
\begin{equation}
\label{B} \mbox{Boundaries:}
\left\{%
\begin{array}{ll}
    0 \rightarrow A & \hbox{\mbox{with rate} $\alpha_{l(r)A}$} \\
    0 \rightarrow B & \hbox{\mbox{with rate} $\alpha_{l(r)B}$} \\
    A \rightarrow 0 & \hbox{\mbox{with rate} $\beta_{l(r)A}$} \\
    B \rightarrow 0 & \hbox{\mbox{with rate} $\beta_{l(r)B}$} \\
    A \rightarrow B & \hbox{\mbox{with rate} $\gamma_{l(r)A}$} \\
    B \rightarrow A & \hbox{\mbox{with rate} $\gamma_{l(r)B}$} \\
\end{array}%
\right.
\end{equation}
in which the indexes $l$ and $r$ indicate the left and the right
boundaries respectively. In the following we will investigate the
steady state properties of this model using the MPF. \\
We assign the operators $E$, $A$ and $B$ to an empty site, an $A$
particle and a $B$ particle respectively. By applying the standard
MPF \cite{dehp} we find the following quadratic algebra for our
model defined by (\ref{A}) and (\ref{B})
\begin{eqnarray}
\label{A2}
& & AE-xEA=\bar{a}E+(\bar{a}+\bar{b})A  \label{A2a} \\
& & BE-xEB=\bar{b}E+(\bar{a}+\bar{b})B  \label{A2b} \\
& & y(AB-BA)=\bar{a}B-\bar{b}A \label{A2c}
\end{eqnarray}
for the bulk operators and also
\begin{eqnarray}
& & [(\beta_{rA}+\gamma_{rA})A-\alpha_{rA}E-\gamma_{rB}B-\bar{a}] \vert V \rangle=0 \label{BA2a} \\
& & [(\beta_{rA}+\gamma_{rB})B-\alpha_{rB}E-\gamma_{rA}B-\bar{b}]\vert V \rangle =0 \label{BA2b} \\
& & \langle W \vert [(\beta_{lA}+\gamma_{lA})A-\alpha_{lA}E-\gamma_{lB}B+\bar{a}]=0 \label{BA2c} \\
& & \langle W \vert [(\beta_{lB}+\gamma_{lB})B-\alpha_{lB}E-\gamma_{lA}A+\bar{b}]=0 \label{BA2d}
\end{eqnarray}
for the boundary terms in which $\bar{a}$ and $\bar{b}$ are
non-zero arbitrary numbers. We have looked for finite-dimensional
representations of the algebra. We have found that a
one-dimensional representation for the algebra exists in which the
operators $A$, $B$ and $E$ are replaced by non-zero real number
$a$, $b$ and $e$. This representation is associated with a uniform
density for each type of particles on the lattice. Defining these
densities as $\rho_A:=\frac{a}{a+b+e}$ and
$\rho_B:=\frac{b}{a+b+e}$ the necessary conditions for the
existence of a one-dimensional representation are
\begin{eqnarray}
\bar{\rho}_A & = & \alpha_{lA}(1-\rho_A-\rho_B)-(\beta_{lA}+\gamma_{lA})\rho_A+\gamma_{lB}\rho_B \label{ODRa}\\
             & = & (\beta_{rA}+\gamma_{rA})\rho_A-\alpha_{rA}(1-\rho_A-\rho_B)-\gamma_{rB}\rho_B \label{ODRb}\\
             & = & (1-x) (1-\rho_A-\rho_B) \rho_A \label{ODRc}\\
\bar{\rho}_B & = & \alpha_{lB}(1-\rho_A-\rho_B)-(\beta_{lB}+\gamma_{lB})\rho_B+\gamma_{lA}\rho_A \label{ODRd}\\
             & = & (\beta_{rB}+\gamma_{rB})\rho_B-\alpha_{rB}(1-\rho_A-\rho_B)-\gamma_{rA}\rho_A \label{ODRe}\\
             & = & (1-x) (1-\rho_A-\rho_B) \rho_B \label{ODRf}
\end{eqnarray}
in which $\bar{\rho}_A:=\frac{\bar{a}}{a+b+e}$ and
$\bar{\rho}_B:=\frac{\bar{b}}{a+b+e}$. We have also found that the
algebra (\ref{A2a}-\ref{BA2d}) has a finite-dimensional
representations provided that the matrices $A$ and $B$ commute
with each other but not with $E$. This means that in the steady
state the probability for finding configurations consisting of
blocks of mixtures of $A$ and $B$ particles surrounded by holes
does not depend on the exact position of the particles in each
block, instead only the number of them in each block will be
important. For instance one can consider the case $A=r \; B$ in
which $r$ is a non-zero real number. This condition indicates that
the density of $A$ particles on each site is always $r$ times the
density of $B$ particles on the same site. By defining a new
operator $D:=A+B=(1+\frac{1}{r})A$ associated with the total
density of particles on each site of the lattice and choosing
$\bar{a}:=\frac{r(1-x)}{1+r}$ it can easily be verified that
(\ref{A2a}) and (\ref{A2b}) both converge to (\ref{A1a}) while
(\ref{A2c}) gives $\bar{b}=\frac{1-x}{1+r}$. By defining two new
parameters $\delta$ and $\beta$ and imposing the following
constraints on the right boundary rates
\begin{eqnarray}
\delta & := & \frac{1}{1-x}(\alpha_{rA}+\alpha_{rB}) = \frac{(1+r)}{r(1-x)} \alpha_{rA} \label{RBCa}  \\
\beta  & := & \frac{1}{1-x}(\beta_{rA}+\gamma_{rA}-\frac{\gamma_{rB}}{r}) = \frac{1}{1-x}(\beta_{rB}+\gamma_{rB}-r\gamma_{rA}) \label{RBCb}
\end{eqnarray}
in which $\alpha_{rA}=r \; \alpha_{rB}$ one can see that the equations
(\ref{BA2a}) and (\ref{BA2b}) become identical to (\ref{A1b}). On the
other hand one can define two new parameters $\alpha$ and $\gamma$ and
impose some constraints on the left boundary rates
\begin{eqnarray}
\alpha & := & \frac{1}{1-x}(\alpha_{lA}+\alpha_{lB}) = \frac{(1+r)}{r(1-x)} \alpha_{lA} \label{LBCa} \\
\gamma & := & \frac{1}{1-x}(\beta_{lA}+\gamma_{lA}-\frac{\gamma_{lB}}{r}) = \frac{1}{1-x}(\beta_{lB}+\gamma_{lB}-r\gamma_{lA}) \label{LBCb}
\end{eqnarray}
in which $\alpha_{lA}=r \; \alpha_{lB}$ to see that the equations
(\ref{BA2c}) and (\ref{BA2d}) become identical to (\ref{A1c}).
Therefore, the operator $D$ associated with the total density of
particles of kind $A$ and $B$ on each site of the lattice besides
the operator $E$ satisfy the PASEP quadratic algebra, provided
that the constraints (\ref{RBCa}-\ref{LBCb}) are satisfied. 
Now one can easily see that on an
special manifold defined by the aforementioned constraints in the
parameters space of our model an $n$-dimensional representation
exists for the quadratic algebra (\ref{A2a}-\ref{BA2d})
provided that (\ref{C1}) is also held. To conclude, we have shown
that under some conditions the quadratic algebra of our
two-species model defined by (\ref{A}) and (\ref{B}) has exactly
one $n$-dimensional irreducible representation for any finite $n$.
Having a finite-dimensional representation for the algebra one can
easily calculate the physical quantities such as the mean density
of particles at each site.
\section{Dynamics of Multiple shocks}
\label{Sec4}
Now we investigate the shock formation and also the shock dynamics
in our two-species model. Since the quadratic algebra of the model
in terms of the total density of particles operator $D$ is exactly
the one for the PASEP and therefore, has a finite-dimensional
representation, one can conclude that multiple shocks might evolve
in the system provided that the constraints (\ref{C2}), (\ref{C3})
and (\ref{C4}) obtained for the PASEP also hold for the total
density of particles in our model defined as
$\rho:=\rho_A+\rho_B$. However, since the density of $A$ particles
on each site is always $r$ times the density of $B$ particles at
the same site it turns out that the multiple shocks structures
exist for both densities $\rho_A$ and $\rho_B$ separately.\\
In the following we will study the case where there is only a
single shock in the system in details. This will be associated with the 
existence of a two-dimensional representation for the quadratic algebra. 
We define the total density of particles on the left- (right-) hand site of the shock position
as $\rho_{l(r)}:=\rho_{l(r)A}+\rho_{l(r)B}$. Since $A=r \; B$ we 
always have
\begin{eqnarray}
\label{C6}
& & \frac{\rho_{lA}}{\rho_{lB}}=\frac{\rho_{rA}}{\rho_{rB}}=r \\
& & \rho_{l(r)A}=\frac{r}{1+r}\rho_{l(r)}\\
& & \rho_{l(r)B}=\frac{1}{1+r}\rho_{l(r)}.
\end{eqnarray}
From our discussion in the previous section and using (\ref{C5}) we
see that in order to have a single shock the total density of
particles on different sides of the shock should satisfy the
condition
\begin{equation}
\label{C7}
\frac{\rho_r(1-\rho_l)}{\rho_l(1-\rho_r)}=\frac{1}{x}.
\end{equation}
On the other hand the total density of particles on each side of
the shock should be obtained from (\ref{C2}) and (\ref{C3})
\begin{eqnarray}
\label{C8}
& & \rho_l = \frac{1}{1+\kappa_{+}(\alpha,\gamma)} \label{ld} \\
& & \rho_r = \frac{\kappa_{+}(\beta,\delta)}{1+\kappa_{+}(\beta,\delta)} \label{rd} 
\end{eqnarray}
in which $\alpha$, $\beta$, $\gamma$ and $\delta$ should be
replaced from (\ref{RBCa}-\ref{LBCb}). It is not difficult to
verify that using (\ref{RBCa}-\ref{LBCb}) the condition
(\ref{ld}) can also be written as
\begin{eqnarray}
\label{C9}
& & \alpha_{lA} = \rho_{lA} ((1-x)+\frac{\beta_{lA}+\gamma_{lA}-\frac{\gamma_{lB}}{r}}{1-\rho_{lA}-\rho_{lB}}) \\
& & \alpha_{lB} = \rho_{lB} ((1-x)+\frac{\beta_{lB}+\gamma_{lB}-r\gamma_{lA}}{1-\rho_{lA}-\rho_{lB}}) \nonumber%
\end{eqnarray}
and (\ref{rd}) as
\begin{eqnarray}
\label{C10}
& & \alpha_{rA} = \rho_{rA} (-(1-x)+\frac{\beta_{rA}+\gamma_{rA}-\frac{\gamma_{rB}}{r}}{1-\rho_{rA}-\rho_{rB}}) \\
& & \alpha_{rB} = \rho_{rB} (-(1-x)+\frac{\beta_{rB}+\gamma_{rB}-r\gamma_{rA}}{1-\rho_{rA}-\rho_{rB}}). \nonumber
\end{eqnarray}
As we mentioned in section \ref{Sec2} the shock position hops to
the left and to the right with the following rates:
\begin{eqnarray}
& & L = \frac{1-\rho_l}{1-\rho_r}x \\
& & R = \frac{1-\rho_r}{1-\rho_l}.
\end{eqnarray}
It is also known that for the PASEP with a single shock the shock
position reflects from the left boundary with rate
$\bar{L}=(1-x)(\frac{\delta}{\rho_r}+\rho_l)$ and also from the
right boundary with rate
$\bar{R}=(1-x)(\frac{\alpha}{\rho_l}-\rho_r)$. For our two-species
model they become
\begin{eqnarray}
& & \bar{L}=\frac{\alpha_{rA}+\alpha_{rB}}{\rho_r}+(1-x)\rho_l \\
& & \bar{R}=\frac{\alpha_{lA}+\alpha_{lB}}{\rho_l}-(1-x)\rho_r.
\end{eqnarray}
In \cite{rs} the authors have introduced a two-species model with
open boundaries similar to the one studied here and investigated
the time evolution of a single shock in the system by introducing
a product shock measure. They have found that under exactly the
same conditions which we introduced in (\ref{RBCa})-(\ref{C10}) 
a stable single shock evolves in the
system which performs a random walk on the lattice. However, in
this paper we have proven that not only one, but also multiple
stable shocks might evolve in the system under slightly different
conditions. The expressions which we have obtained for the hopping
rate of the shock positions in the bulk of the lattice and also
the reflection rates from the boundaries are quit in agreement
with those obtained in \cite{rs}. One can easily check that the
necessary conditions for having a stationary product measure with
uniform densities for both kinds of particles obtained in
\cite{rs} are exactly the conditions for the existence
of a one-dimensional representation of the quadratic algebra
(\ref{A2a}-\ref{BA2d}) obtained in (\ref{ODRa}-\ref{ODRf}).
\section{Concluding remarks}
\label{Sec5}
In this paper we have studied a two-species model with open
boundaries and the conditions under which an
invariant multiple shocks measure exists for it. We have found that from
the MPF point of view the quadratic algebra of the model can be
mapped to that obtained for the PASEP, if one defines the total
density of particles, and also redefines the boundary conditions
(the sum of the densities of particles at each site is defined as
the total density of particles at that site). Then we have been
able to introduce a finite-dimensional Fock representation for
this quadratic algebra, provided that the microscopic rates satisfy
some constraints. It terms of the total density of particles these
constraints are nothing but the necessary conditions for the
existence of multiple shocks in the system. The associated
quadratic algebra of the model has also a one-dimensional
representation under some restrictions which are the conditions
for the existence of a uniform stationary product measure for the
model. For the single-shock case our results obtained from MPF
analysis agree with those obtained in \cite{rs} from study of
the product shock measure dynamics. \\
In \cite{kjs} the authors have shown that for three different
families of one-dimensional driven diffusive models with open
boundaries, the stationary state distribution function can be
written as a linear superposition of single shock measures,
provided that some constraints are fulfilled. In \cite{far} the
author has studied the same models and shown that from the MPF point
of view these constraints are the necessary conditions for the
existence of two-dimensional representations for their quadratic
algebras. The problem now worth studying is whether
every finite-dimensional representation of the quadratic
algebra of a given one-dimensional driven diffusive model with
open boundaries is associated with expression of its stationary
state distribution function in terms of superposition of product 
shock measures. This is still under
investigation.

\end{document}